\documentstyle[epsfig]{aipproc}

\begin{document}

\newcommand{\nub}{\mbox{$\overline{\nu}$}}
\newcommand{\num}{\mbox{$\nu_\mu$}}
\newcommand{\numb}{\mbox{$\overline{\nu_\mu}$}}
\newcommand{\nue}{\mbox{$\nu_e$}}
\newcommand{\nueb}{\mbox{$\overline{\nu_e}$}}
\newcommand{\rmt}{\rm\textstyle}

\title{A Small Target Neutrino Deep-Inelastic Scattering 
Experiment at the First Muon Collider}

\author{Deborah A. Harris$^*$ and Kevin S. McFarland$^{\dagger}$}
\address{$^*$University of Rochester, Rochester, NY 14627 \\
 $^{\dagger}$Massachusetts Institute of Technology, 
77 Massachusetts Ave., Cambridge, MA 02139}
\maketitle
\vspace{-.25in} 
\begin{abstract}
Several different scenarios for neutrino scattering experiments 
using a neutrino beam from the muon collider complex are 
discussed.  The physics reach of a neutrino experiment
at the front end of a muon collider is shown to extend far 
beyond that of current neutrino experiments, 
since the high intensity neutrino beams one
would see at the muon collider allow for a large 
flexibility in choosing neutrino targets.  Measurements of quark
spin, A-dependence of the structure function $xF_3$ and neutral current
chiral couplings to quarks are outlined.  
\end{abstract}
\mbox{\parbox{\textwidth}{
\vspace{-9.in} 
\begin{flushright}
LNS-98-277\\ UR-1516 ER$/$4056$/$911 \\ FNAL-CONF-98/119
\end{flushright} 
}}
\vspace{-.75in} 
\section*{Introduction} 
     Neutrino deep-inelastic scattering has proven an invaluable tool 
to study hadron structure, QCD, and the electroweak force.  
Neutrinos give the only clean measurement of the valence quark 
distribution inside a nucleon, and have a very distinct signature 
for scattering off a strange quark inside the nucleon sea.  
Neutrino-nucleon scattering also provides two important  
tests of QCD, through structure function evolution and the 
Gross Llewellyn Smith Sum Rule.  Both tests can also provide 
precise measurements of the strong coupling constant, 
$\alpha_s$ \cite{alfs-meas}.  Neutrino scattering experiments
which can reconstruct neutral current interactions also provide 
a stringent test of electroweak symmetry breaking.  

     In the past, neutrino experiments have been shaped by the low 
neutrino scattering cross section.  To get ample statistics 
experiments have had to use massive active targets.  For 
example, the CCFR/NuTeV experiments used  a target of 700 tons of steel instrumented
with scintillators\cite{billsf}, and the NOMAD experiment used a target of  
$2.7$ tons of drift chambers\cite{nomad}.  In reference \cite{other_proc} the 
results from a simulation of the neutrino beams resulting from 
high energy muon beams were given, and event 
samples of comparable statistics are attainable using targets of only 
several $g/cm^2$ thickness.  In this 
paper we consider targets that have previously been used 
for muon scattering, and discuss the novel physics that suddenly 
becomes feasible when placing these targets in extremely intense 
neutrino beams.  

\section*{Neutrino Scattering Kinematics} 

\begin{figure}[h!] 
\epsfxsize=\textwidth\epsfbox{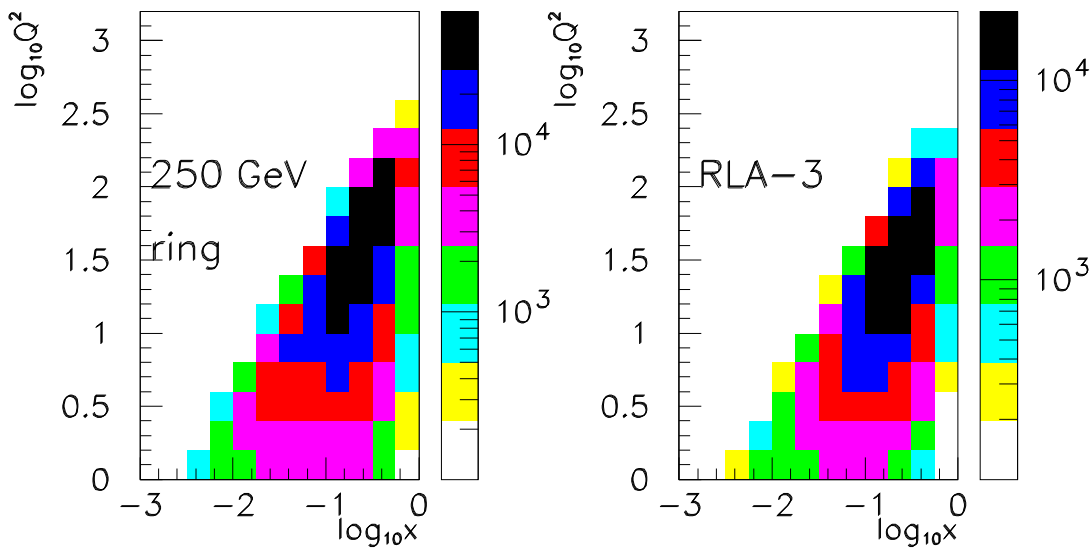}
\vspace{10pt} 
\caption{$x$ versus $Q^2$ distributions for 
two possible neutrino experiments.  Units are in events per year per 
$g/cm^2$ of target for a $20\,cm$ radius target.} 
\label{fig:xq2} 
\end{figure}

By including known $\nu$-nucleon 
differential cross sections in the Monte Carlo
described in reference \cite{other_proc}, one can 
predict the kinematic reach of a year's worth of data per $g/cm^2$ 
of target.  It is important to note that since the 
neutrino beams from a muon collider are not more energetic than
conventional beams, the reach in momentum transfer squared ($Q^2$) 
of these experiments is comparable 
to what has already been measured.  However, since $\alpha_s$ is large 
and changing rapidly in the region $1<Q^2<100\,GeV^2$, this is precisely 
where one wants to make perturbative QCD measurements.  The kinematic variable 
one could extend in reach is $x$, or the struck quark's fractional 
momentum.  To probe smaller and larger $x$ regions one needs 
both high statistics and good detector acceptance and resolution.  
To maximize statistics one could use a target 
which would result in a optimal number of interactions
(for example, $.05$) per 
turn.  This target would still be rather small, and would not 
degrade the event reconstruction, compared to previous neutrino 
experiments.  The remainder of the experiment could then be a low-mass 
spectrometer with some particle identification, 
modeled on non-neutrino fixed target experiments 
(see reference \cite{other_proc}).  A target of 
the size described above with a low mass detector to 
reconstruct the scatters would result in about $2\times10^7$ events per year 
for one of the recirculating linacs, and a factor of 100 more
than that for an experiment downstream of a collider ring's 
straight section.    
Figure \ref{fig:xq2} shows the number of events expected
per $g/cm^2$ as a function of $x$ and $Q^2$, assuming a 
$20\,cm$ radius target.  The reach in $x$ for a ``high statistics target'' 
in a given $Q^2$ region is an order of magnitude beyond what is 
currently measured.  
As there have been surprises in the low $x$ region, for example, 
the fast rise of $F_2$ as seen by HERA \cite{heralowx},   
the question arises, is the valence quark distribution 
well-behaved at low $x$, or does something unexpected happen there too?   

\subsection*{Strong Coupling Constant Determinations} 

Neutrino deep-inelastic scattering provides two clean measurements of 
the strong coupling constant $\alpha_s$: one from the $Q^2$ dependence
of the non-singlet structure function $xF_3$, and one from the Gross
Llewellyn Smith Sum Rule.  Both of these measurements are independent
of the gluon distribution, and the latter has very small perturbative QCD 
uncertainties.  Currently, however, both measurements are limited by 
experimental systematic uncertainties:  the $xF_3$ evolution is 
limited by the uncertainty in the energy calibration \cite{billsf}, 
and the GLS Sum 
Rule is limited by the uncertainty in both the overall neutrino 
cross section, 
and the ratio of neutrino to anti-neutrino cross sections \cite{GLS}.  
Furthermore, 
both cross section measurements are themselves limited by systematic 
uncertainties \cite{billthesis}.  A neutrino experiment at a muon collider
could address all three of these issues in fundamentally different ways, 
because of the way the beam is formed.  

Neutrinos from a monochromatic beam of muons have a very distinct 
spectrum peaked towards the high end of the available energy from 
the muon.  Muon decay itself is very well understood, and the muon energies
at the collider are predicted to be measurable to a few parts per million
\cite{alvin-raja}.  By calibrating the 
detector to the end point of the neutrino spectrum one should be 
able to achieve much better than the current energy scale uncertainty 
in neutrino experiments, which is about $1\%$\cite{billsf}.  

Similarly, the muon currents in the beam-lines can be measured, and 
given the energy of the muons and the currents and the geometry of 
the beam-line, the neutrino flux at a given detector can be accurately  
predicted.  The availability of an independent flux measurement for a 
neutrino experiment will be an important improvement and will dramatically
change the nature of absolute neutrino cross section measurements, 
as well as the scale of the uncertainties in those measurements.  

\section*{Light Targets}  

Large low-Z targets have already been used successfully 
by muon scattering experiments.  The sizes of liquid hydrogen 
targets are in general limited by safety considerations, and a liquid
hydrogen 
target of $1m$ thickness and radius approximately $10\,cm$ has been used 
by experiment E665 at Fermilab \cite{pgs}.  A target of this volume
of hydrogen with a $20\,cm$ radius would have a thickness of 
$1.75 g/cm^2$, resulting in data samples of slightly over a million 
$\nu$ charged current events per year.  The per turn event rate on 
this target would still be small enough so that pile-up in the target
itself would not be a problem.

\section*{Nuclear Effects}  
     Charged lepton experiments have studied nuclear dependences of 
the structure function $F_2$ and have seen a sizeable effect 
\cite{emc}\cite{pgs}, 
yet no nuclear dependences have been seen in $xF_3$, and for now theory must
only assume that the integral of the valence quark distribution is independent 
of nuclear corrections.  By scattering neutrinos off several light targets
one could finally determine whether or 
not valence quarks know anything about what kind of nucleus they are in.  
If one had both 
a hydrogen and a deuterium target, one could measure if $u(x)$ in the 
proton is indeed equal to $d(x)$ in the neutron.  Such a measurement would 
either confirm or challenge our assumptions about isospin symmetry.  
The effects seen in charged lepton scattering are large; for example, 
$2F_2^d/F_2^p-1$ is $0.935\pm0.008 ({\rm stat})\pm0.034({\rm syst})$ 
at $x$ below $0.01$
\cite{pgs}, and at higher $x$ this ratio is even farther away from 
unity.  With hydrogen and deuterium   
targets of the sizes suggested above one would have enough statistics
to measure the quantity $2xF_3^d/xF_3^p-1$ as a function of $x$ to 
better than a per cent, over a reasonably large $x$ range.  

\section*{Polarized Targets}  

\begin{figure}[tbp]
\hspace*{0.15\textwidth}\epsfxsize=0.7\textwidth
\epsfbox[100 220 485 540]{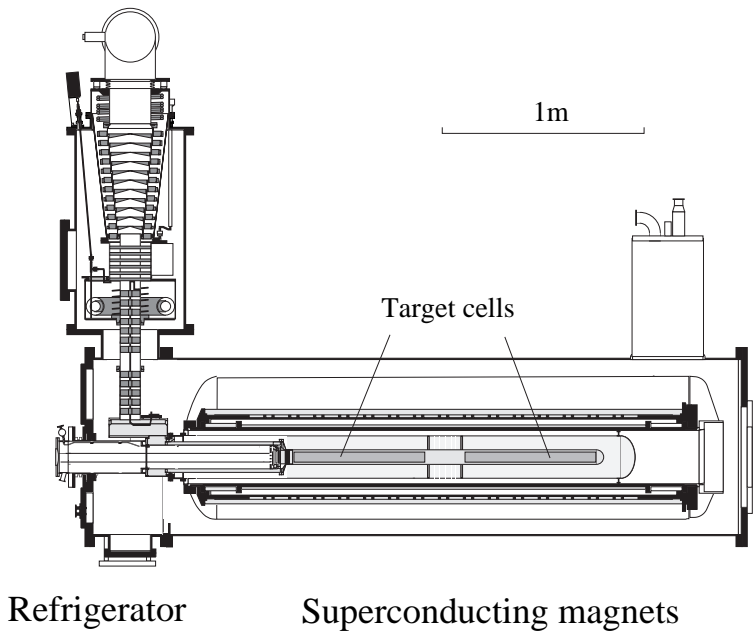}
\caption{Polarized butanol target used in the SMC experiment (see text).
The total target mass was $3$~kg, and the polarized fraction of the target was
approximately $10\%$.}
\label{fig:smc-targ}
\end{figure}
One exciting possibility in such a light-target neutrino experiment is
measuring neutrino scattering from polarized targets.
The neutrino beams from the muon collider can produce on the 
order of $5\times10^5$ events/kg/year
in a target; therefore, a target with approximately $20$~kg of polarized
protons would produce excellent statistics.

An example of a large polarized target used in the past is the SMC solid 
butanol target\cite{smc-targ} (Figure~\ref{fig:smc-targ}), 
which was a $2.5$~cm radius cylinder containing $0.3$~kg of polarized protons.
If such a target could be scaled up to the size of the neutrino beam,
approximately $20$~cm in radius, this would provide sufficient mass for a
high statistics neutrino experiment.

\section*{Spin Physics} 

If the polarized target described above were feasible, the
physics motivations would be many-fold.  Deep-inelastic scattering from
hydrogen targets would probe the following processes
$$
\begin{array}{cc}
\nu u\to\mu^-d & \nu \overline{d}\to\mu^-\overline{u} \\
\nub d\to\mu^+u & \nub \overline{u}\to\mu^+\overline{d} \\
\nub s\to\mu^+c & \nu \overline{s}\to\mu^-\overline{c},
\end{array}
$$
and their Cabbibo-suppressed analogs.  The beauty of probing polarized
targets with neutrinos is that these six processes are clearly separable.
$y$ distributions can be used to separate quark and anti-quark targets;
neutrinos and anti-neutrinos each pick one flavor of a quark doublet;
strange quarks can be tagged by final state charm in its decay to high momentum
leptons or {\em via} its finite lifetime.

This leads to the possibility of a measurement of the polarization of the
proton's quarks by flavor, with sea and valence contributions separated.
This measurement would be relatively independent of details of final
state hadronization in contrast with current experimental programs of
measuring the polarization of the strange sea which rely on 
final state strange mesons or baryons.

The statistical challenge for such an experiment would be the
measurement of the strange polarization from final state di-lepton
events, where one lepton comes from the decay of charm.  The total
di-lepton cross-section is approximately $2\%$, therefore, on a $200$~kg
butanol target, approximately $1\times10^6$ neutrino and $5\times10^5$
anti-neutrino di-lepton events per year would be observed, about half of
which come from scattering off of strange quarks.  With these statistics,
a raw asymmetry could be measured with a $1.5\times10^{-3}$ precision.
Assuming the $d$ quark asymmetry would be better known from the more
common processes listed above, this would translate into a precision
in the strange sea polarization of $3\%$ in a one year run.

This precision is reasonable for testing models for the spin of the
strange sea.  A recent prediction for the strange sea polarization in
an effective chiral quark model\cite{Quigg} estimates a $-8\%$
polarization.

\section*{Neutral Current Quark Couplings} 

Neutrino deep-inelastic scattering from light, non-isoscalar
nuclei also gives the possibility of doing precision measurements of
the left- and right-handed couplings of the neutral current to quarks.
To date, precision neutrino and atomic physics experiments have probed
primarily heavy nuclei and thus could only measure isoscalar combinations
of couplings.  Such observations could determine whether the lingering
evidence for some enhanced coupling of the neutral current to $b$-type
quarks \cite{pdg-summ} is only observed in the $b$ system or whether it 
is perhaps the first evidence of some generation-independent phenomenon.

\typeout{do we want some effort to do numbers?}
 
\section*{Conclusions} 

We have presented several different physics motivations for a 
deep-inelastic scattering neutrino experiment at the front end of the 
muon collider complex.  The possibility of scattering neutrinos 
off either light elements or polarized targets 
will address a wide range of issues in our understanding of hadron
structure.  The experience from polarized targets used in muon 
scattering experiments will be particularly useful, as it would 
enable measurements of the individual quark contributions to the
spin of the proton.  In order to best take advantage of the neutrino
beams at the muon collider complex a run of two or more years using 
several different cryogenic targets would be optimal.  Furthermore, 
the characteristics of this new neutrino beam would be 
ideal for new precision measurements of both the strong coupling
constant, and the neutral current quark couplings.  In summary, 
deep-inelastic scattering neutrino experiments at the 
muon collider have the potential to fundamentally change our understanding
of the baryons that make up most of the observed universe.

\end{document}